\documentclass[10pt,amsmath,amssymb,aps,nofootinbib,prd,twocolumn]{revtex4-1}
\usepackage{aas_macros,ae,aecompl,graphicx,hyperref,subfigure}
\usepackage[pass]{geometry}
\usepackage[T1]{fontenc}

\renewcommand{\O}{\Omega}

\begin{document}

\title{Note on Bunching of Field Lines in Black Hole Magnetospheres}

\author{\bf Samuel E. Gralla}
\email{sgralla@physics.harvard.edu}
\author{\bf Alexandru Lupsasca}
\email{lupsasca@fas.harvard.edu}
\author{\bf Maria J. Rodriguez\,}
\email{mjrodri@physics.harvard.edu}

\affiliation{
	\vskip 0.5mm
	{\it Center for the Fundamental Laws of Nature, Harvard University, Cambridge 02138, MA USA}
	\vskip 0.5mm
}

\begin{abstract}
	Numerical simulations of Blandford-Znajek energy extraction at high spin have revealed that field lines tend to bunch near the poles of the event horizon. We show that this behavior can be derived analytically from the assumption of fixed functional dependence of current and field line rotation on magnetic flux. The argument relies crucially on the existence of the Znajek condition, which offers non-trivial information about the fields on the horizon without requiring a full force-free solution. We also provide some new analytic expressions for the parabolic field configuration.
\end{abstract}

\maketitle

%%%%%%%%%%%%%%%%%%%%%%%%%%%%%%%%%%%%%%%%
\section{Introduction}

A leading candidate mechanism to power relativistic jets from active galactic nuclei (AGN) is the Blandford-Znajek (BZ) process \cite{Blandford:1977ds}, in which energy is extracted from a spinning black hole via its plasma magnetosphere. In light of the large observed variation in jet properties, it is of interest to explore the detailed dependence of magnetosphere structure on system parameters. One particular variation was noticed by \cite{Tchekhovskoy:2009ba} (TNM), who found that as the black hole spin is increased, the magnetic flux on the event horizon tends to concentrate near the pole, a feature they described as {\it bunching} of field lines. In this note we will show how the bunching can be derived by a simple analytic argument from an assumption about the scaling of the current and field line rotation with spin. The argument makes no reference to a particular choice of field line geometry, and we thereby show that the bunching is a general phenomenon, not limited to the particular geometries considered by TNM. We illustrate the high-spin bunching for the three cases where approximate analytical solutions are known, the radial \cite{Michel:1973,Blandford:1977ds}, parabolic \cite{Blandford:1976,Blandford:1977ds}, and hyperbolic \cite{Beskin:1992,Gralla:2015vta} field configurations.

%%%%%%%%%%%%%%%%%%%%%%%%%%%%%%%%%%%%%%%%
\section{General Magnetosphere Structure}

Following BZ and TNM we assume that the plasma inertia is negligible, so that the magnetosphere is force-free.\footnote{Reviews of force-free electrodynamics, including the discussion of the quantities relevant to this work, may be found in \cite{Beskin:2010iba,Gralla:2014yja}.} A stationary, axisymmetric force-free field in a spinning black hole background is fully characterized by the flux function $\psi(r,\theta)$, the polar current $I(\psi)$, and the angular velocity of field lines $\O(\psi)$.\footnote{We will use the conventions of TNM.} These functions encode the energy flux by 
\begin{align}
	\frac{d\mathcal{E}}{dt}=2\int_{0}^{\psi_*}\!d\psi\,I(\psi)\,\O(\psi),
\end{align}
where we assume a reflection-symmetric magnetosphere and define $\psi$ on the northern hemisphere, with $\psi$ zero at the pole and monotonically increasing until it reaches its largest value $\psi_*$ on the equator. (The factor of $2$ accounts for the southern hemisphere.) Three families of energy-extracting approximate solutions are known (radial, parabolic, and hyperbolic) and in all cases the current and angular velocity take the form
\begin{align}
\label{eq:assumption}
	I(\psi)=\O_H\,\mathcal{I}(\psi),\qquad\O=\O_H\,\mathcal{O}(\psi),
\end{align}
where $\O_H$ is the horizon angular velocity and $\mathcal{I}$ and $\mathcal{O}$ do not depend on the spin $a$. This leads to the basic prediction of the BZ model that the power scales as the spin squared,
\begin{align}
	\frac{d\mathcal{E}}{dt}\propto\O_H^2.
\end{align}
This result was derived analytically for small spin, but TNM has shown that for generic field configurations, it continues to hold for all but the highest ($a\gtrsim0.99M$) spins.\footnote{As TNM show, it important to write the linearized result as $\O_H^2$, rather than (say) $a^2$, in order for the scaling to carry over to high spin.}$^{,}$\footnote{To define the notion of the ``same'' field configuration at different spins, TNM use the same initial data at each $(r,\theta)$ in Boyer-Lindquist coordinates.} In fact TNM find more: not just the total power but also the detailed functional forms in Eqs.~\eqref{eq:assumption} carry over from the linearized theory to large spin. In particular, their Figs.~5 and 6 demonstrate that $\mathcal{I}(\psi)$ and $\mathcal{O}(\psi)$ vary by no more than 10\% over the entire range of spins $0<a/M<0.9999$. We will take this result as our starting point, and for the remainder of the paper we will assume that Eqs.~\eqref{eq:assumption} hold \textit{exactly} for any spin, with $\mathcal{I}(\psi)$ and $\mathcal{O}(\psi)$ independent of the spin parameter $a$. We will also assume that the total magnetic flux $\psi_*$ is independent of spin. We now show how the bunching of field lines can be recovered straightforwardly from these assumptions.

%%%%%%%%%%%%%%%%%%%%%%%%%%%%%%%%%%%%%%%%
\subsection{Bunching of field lines}

Even in possession of the detailed functional forms of $I(\psi)$ and $\O(\psi)$, determining a complete force-free solution is a daunting task, requiring the solution of a second-order, non-linear partial differential equation for $\psi$. However, for a Kerr black hole this equation has the remarkable property that, when evaluated on the horizon, only derivatives tangential to the horizon appear, and furthermore the equation can be integrated once, so that it becomes a first-order ordinary differential equation on the horizon. The result is the so-called Znajek condition (\cite{Znajek:1977} and e.g. \cite{Gralla:2014yja}),
\begin{align}
\label{eq:Znajek}
	I=-2\pi(\O-\O_H)\frac{\left(r_H^2+a^2\right)\sin{\theta}}{r_H^2+a^2\cos^2{\theta}}\,\partial_\theta\psi,
\end{align}
where $\O_H=a/\left(r_H^2+a^2\right)$ is the horizon angular velocity, $r_H=M+\sqrt{M^2-a^2}$ is the horizon radius, and everything is evaluated on the horizon. We can integrate and then exponentiate Eq.~\eqref{eq:Znajek} to find
\begin{align}
\label{eq:piggy}
	f(\psi)=A(\theta),
\end{align}
where we defined
\begin{align}
\label{eq:f}
	f(\psi)\equiv\exp\left[-2\pi\int\!d\psi\,\frac{\O-\O_H}{I}\right] 
\end{align}
and
\begin{align}
	A(\theta)\equiv e^{a\,\O_H\cos{\theta}}\tan{\tfrac{\theta}{2}}.
\end{align}
Eq.~\eqref{eq:f} defines $f(\psi)$ up to an overall constant to be fixed, after solving Eq.~\eqref{eq:piggy}, by demanding that the maximum value of $\psi$ be $\psi_*$. Formally, the solution to Eq.~\eqref{eq:piggy} is
\begin{align}
	\psi_H(\theta)=f^{-1}[A(\theta)]=f^{-1}\left[e^{a\,\O_H\cos{\theta}}\tan{\tfrac{\theta}{2}}\right],
\end{align}
where the notation $\psi\equiv\psi_H$ serves as a reminder that this formula holds on the horizon. Given the assumptions that $\mathcal{I}(\psi)$, $\mathcal{O}(\psi)$, and $\psi_*$ are independent of the spin, we know that $f$, and hence $f^{-1}$, is likewise independent of the spin. Thus the {\it only} dependence of the horizon flux on the spin is through the factor $e^{a\,\O_H\cos{\theta}}$ in the argument of $f^{-1}$. This factor is monotonically decreasing from its maximum $e^{a\,\O_H}$ at the pole to its minimum $1$ at the equator, and will therefore always tend to increase the proportion of magnetic flux near the pole as the spin is increased. This is the bunching of field lines.

%%%%%%%%%%%%%%%%%%%%%%%%%%%%%%%%%%%%%%%%
\section{Specific Magnetic Geometries}

We now consider force-free solutions with radial, parabolic, and hyperbolic geometries. In each case we use the current and angular velocity functions appropriate to a normalization of $\psi(\pi/2)=\psi_*=1$. Factors of $\psi_*$ may be reinstated by  scaling $\psi\rightarrow\psi/\psi_*$, $I\rightarrow I/\psi_*$ and $\O\rightarrow\O$. E.g., equation (\ref{eq:radial}) becomes $I(\psi)=2\pi\,\O(\psi)\,\psi(2-\psi/\psi_*)$.

%%%%%%%%%%%%%%%%%%%%%%%%%%%%%%%%%%%%%%%%
\subsection{Radial}

In their original paper BZ found an approximate solution with radial field lines in a ``split monopole'' configuration. For $\psi_*=1$ the current and angular velocity are given by
\begin{align}\label{eq:radial}
	I(\psi)&=2\pi\,\O(\psi)\,\psi(2-\psi),\\
	\O(\psi)&=\tfrac{1}{2}\O_H,
\end{align}
which satisfy our assumptions with $\mathcal{I}=\pi\psi(2-\psi)$ and $\mathcal{O}=1/2$. These are depicted in Fig.~\ref{fig:radial}, which can be matched directly to Fig.~4 of TNM. The integral in Eq.~\eqref{eq:f} becomes
\begin{align}
	f(\psi)=C\sqrt{\frac{\psi}{2-\psi}},
\end{align}
for some constant $C$. We can then solve Eq.~\eqref{eq:piggy} on the horizon to learn that $\psi=2A^2/\left(C^2+A^2\right)$. Requiring $\psi(\pi/2)=\psi_*=1$ then fixes $C^2=1$, and hence
\begin{align}
	\psi_H(\theta)=\frac{2A^2(\theta)}{1+A^2(\theta)}.
\end{align}
To compare directly to TNM we plot the radial magnetic field $B^r=\partial_\theta\psi/\sqrt{-g}$ at the event horizon $r=r_H$ (Fig.~\ref{fig:radial}) where $\sqrt{-g}=\left(r^2+a^2\cos^2{\theta}\right)\sin{\theta}$. Comparing with their Fig.~7, we see excellent agreement for all curves $a\le0.9$, with TNM seeing more bunching for larger spins. This discrepancy can be explained by the fact that our assumption of spin-independent $\mathcal{I}(\psi)$ and $\mathcal{O}(\psi)$ disagrees more with TNM at higher spins.

%%%%%%%%%%%%%%%%%%%%%%%%%%%%%%%%%%%%%%%%
\subsection{Parabolic}

BZ also found an approximate solution with parabolic field lines, whose current and angular velocity are
\begin{align}
	I(\psi)&=4\pi\,\O(\psi)\,\psi,\\
	\label{eq:parabolic}
	\O(\psi)&=\O_H\frac{(1-\psi)(u+1)[u-(1-\psi)\ln2]}{u^2+(1-\psi)\,u-(1-\psi)^2(u+1)\ln2},
\end{align}
where $u=F[(1-\psi)\ln4]$ and $y=F(x)$ is the product logarithm, which is defined by the principal solution of $x=ye^y$.\footnote{The angular velocity $\O$ is normally given in terms of coordinates rather than as a function of $\psi$. As far as we know, Eq.~\eqref{eq:parabolic} is a new expression.} Thus this configuration also has spin-independent $\mathcal{I}(\psi)$ and $\mathcal{O}(\psi)$.

Following the same steps as before, we find that
\begin{align}
	f(\psi)=C\sqrt{\frac{u}{(1-\psi)\ln2}-1},
\end{align}
and for the horizon flux function that
\begin{align}
	\psi_H(\theta)=\frac{A^2(\theta)\ln2+\ln[1+A^2(\theta)]}{\left[1+A^2(\theta)\right]\ln2}.
\end{align}
(Demanding $\psi_H(\pi/2)=\psi_*=1$ has fixed the constant to be $C^2=1$.) The radial magnetic field is plotted in Fig.~\ref{fig:parabolic}, and shows even more pronounced bunching than in the monopolar case.

%%%%%%%%%%%%%%%%%%%%%%%%%%%%%%%%%%%%%%%%
\subsection{Hyperbolic}

\begin{figure*}
\centering
\subfigure[\ radial]{
	\label{fig:radial}
	\includegraphics[width=7cm,height=5cm]{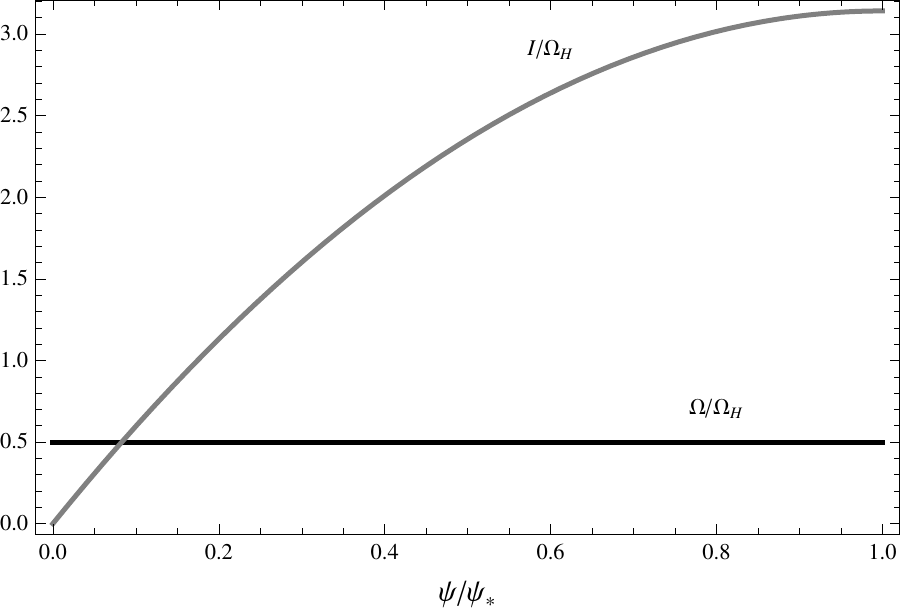}
	\includegraphics[width=7.5cm,height=5.05cm]{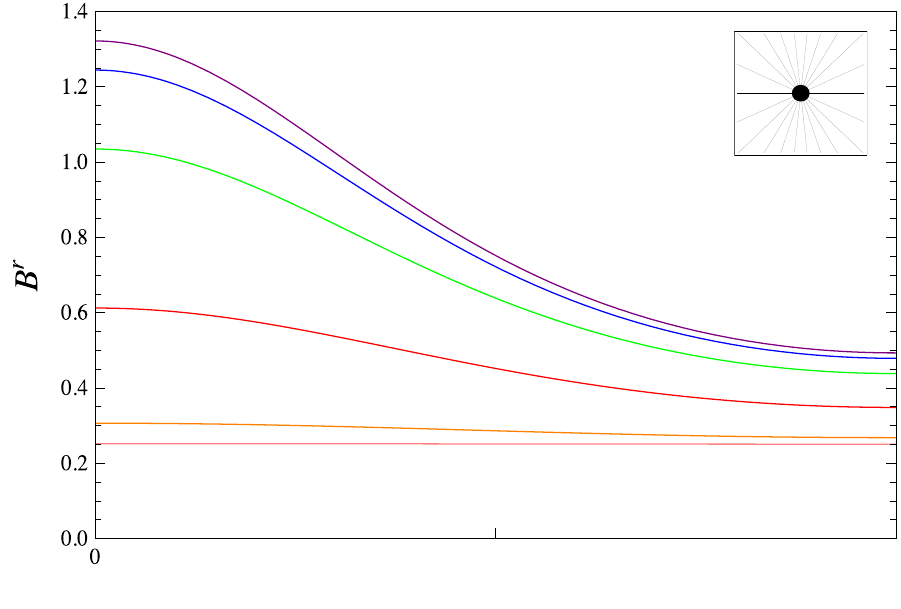}\vspace{50mm}
\begin{picture}(0,0)(0,0)
\put(-10,8){$\pi/2$}
\put(-106,8){$\pi/4$}
\put(-90,-2){$\theta$}
\end{picture}
}\\
\subfigure[\ parabolic]{
	\label{fig:parabolic}
	\includegraphics[width=7cm,height=5cm]{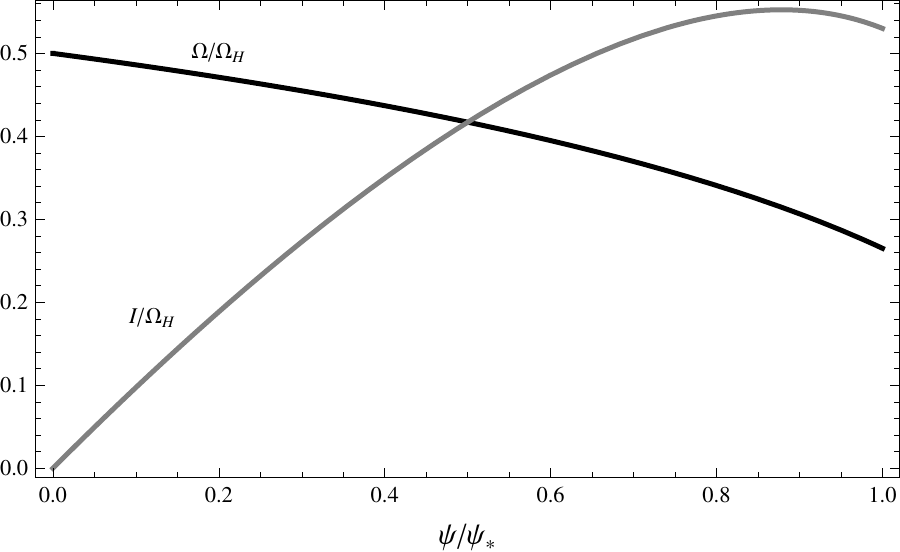}
	\includegraphics[width=7.5cm,height=5.05cm]{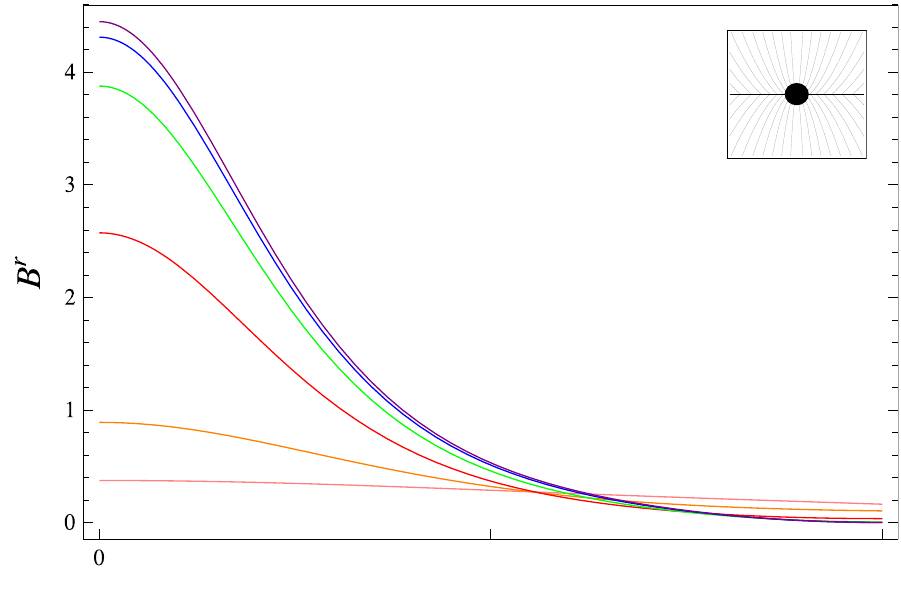}
	\begin{picture}(0,0)(0,0)
\put(-10,8){$\pi/2$}
\put(-106,8){$\pi/4$}
\put(-90,-2){$\theta$}
\end{picture}
}\\
\subfigure[\ hyperbolic]{
	\label{fig:vertical}
	\includegraphics[width=7cm,height=5cm]{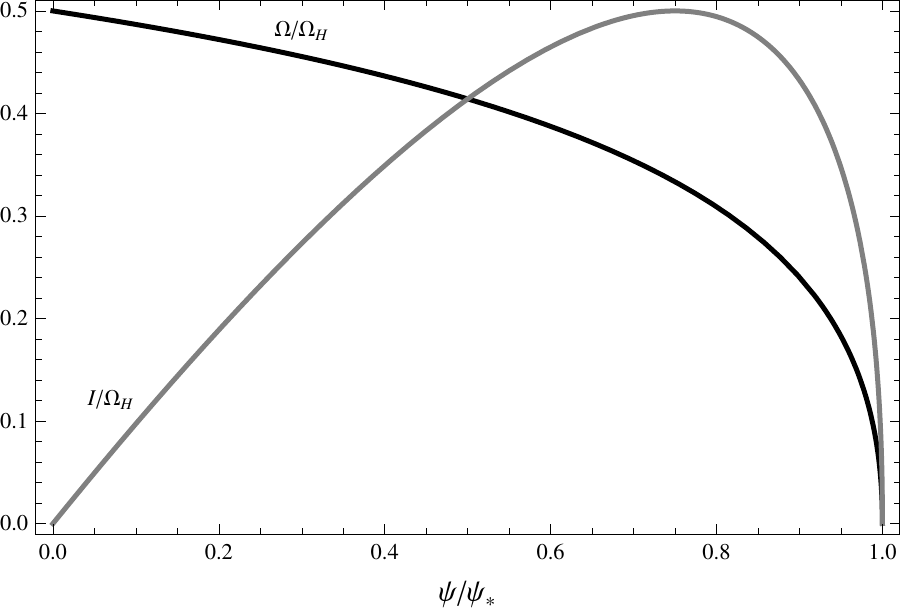}
	\includegraphics[width=7.5cm,height=5.05cm]{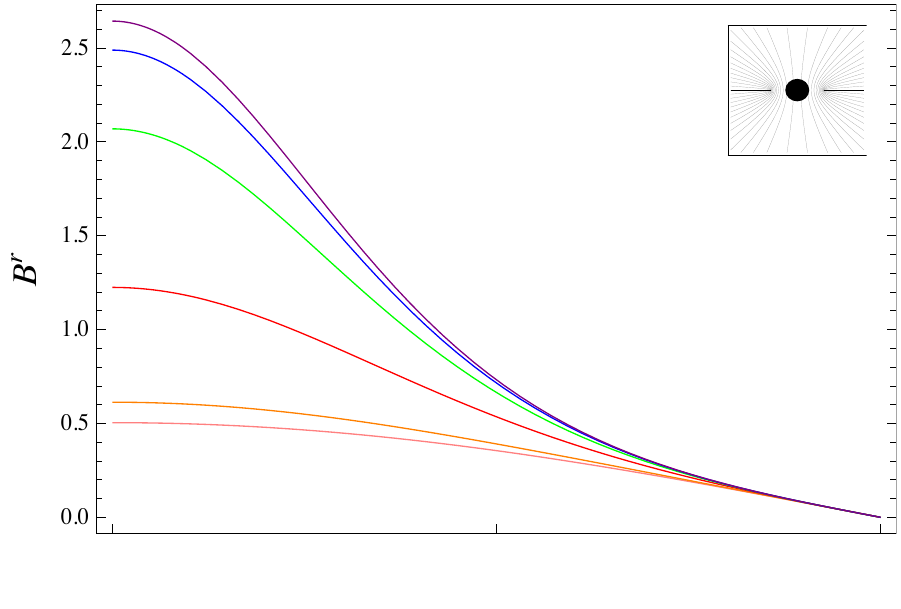}
\begin{picture}(0,0)(0,0)
\put(-10,10){$\pi/2$}
\put(-10,10){$\pi/2$}
\put(-106,10){$\pi/4$}
\put(-90,0){$\theta$}
\put(-192,10){$0$}
\end{picture}
}
\caption{Bunching of field lines for radial, parabolic, and hyperbolic geometries. Diagrams of the field line geometry appear in the upper right. On the left, the dimensionless current $\mathcal{I}$ ({\it black line}) and angular velocity $\mathcal{O}$ ({\it gray line}) as a function of magnetic flux $\psi/\psi_*$ on the horizon $\psi<\psi_*$, which we assume to hold at all spin. On the right, the radial magnetic field $B^r=\partial_{\theta}\psi/\sqrt{-g}$ as a function of angle $\theta<\pi/2$ on the horizon for different values of the spin $a=0.1,0.5,0.9,0.99,0.999,0.9999$ (upwards on the left).}
\label{fig:figure}
\end{figure*}
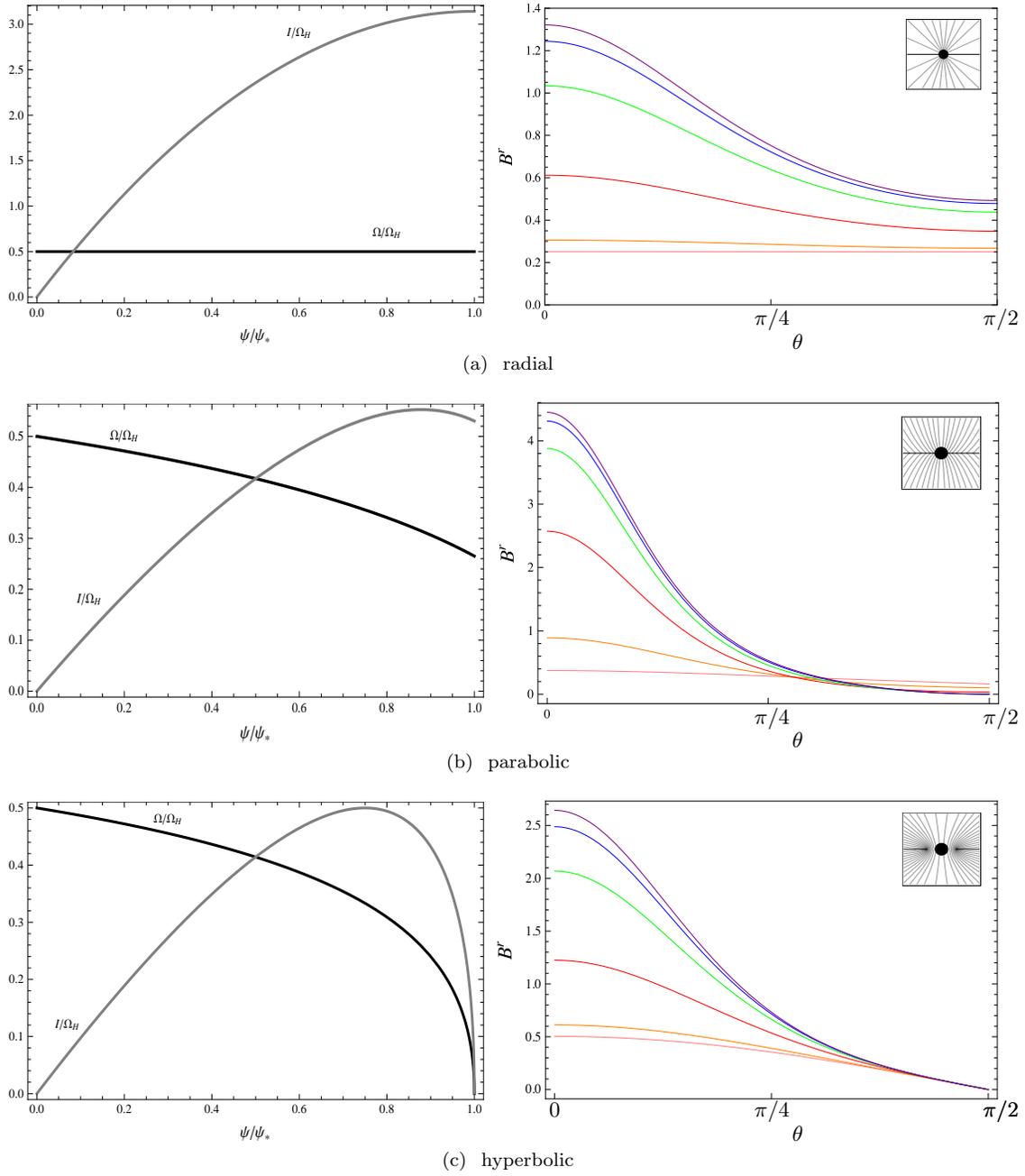

A third, ``hyperbolic'' solution was found in \cite{Beskin:1992,Gralla:2015vta}, where the field is generated by a thin disk terminating at an inner radius $b$. The current and angular velocity functions satisfy our assumptions. For simplicity we will work in the limit $b\gg M$, where they become\footnote{One may also obtain these expressions by considering precisely vertical field lines and demanding the convergence of an integral formula for higher order corrections \cite{Pan:2014bja}.} 
\begin{align}
	I(\psi)&=4\pi\,\O(\psi)\,\psi,\\
	\O(\psi)&=\O_H\frac{\sqrt{1-\psi}}{1+\sqrt{1-\psi}}.
\end{align}
These expressions are a good approximation even when $b$ corresponds to the innermost stable circular orbit. Performing the bunching calculation, we find
\begin{align}
	f(\psi)=C\sqrt{\frac{1-\sqrt{1-\psi}}{1+\sqrt{1-\psi}}},
\end{align}
and
\begin{align}
	\psi_H(\theta)=\frac{4A^2(\theta)}{\left[1+A^2(\theta)\right]^2},
\end{align}
where $\psi_H(\pi/2)=\psi_*=1$ has fixed $C^2=1$. The radial component of the magnetic field at the horizon is plotted in Fig.~\ref{fig:vertical}, which again shows the bunching.

%%%%%%%%%%%%%%%%%%%%%%%%%%%%%%%%%%%%%%%%
\section{Summary}
\label{Discussion}

We have shown that the high-spin bunching of field lines observed by TNM generalizes to arbitrary magnetic geometry under the assumption that the functional forms of the current $I(\psi)$ and field line angular velocity $\O(\psi)$ both scale linearly with $\O_H$. Key to enabling this analytic argument was the existence of the Znajek condition at the horizon. This technique allows us to bypass the issue of having to solve the complete non-linear force-free problem to derive properties of the fields at the horizon. We gave a general argument and then illustrated the bunching with three specific magnetic geometries, which are plotted in Fig.~\ref{fig:figure}.\\

%%%%%%%%%%%%%%%%%%%%%%%%%%%%%%%%%%%%%%%%
\section*{Acknowledgements}

We would like to thank A. Tchekhovskoy for useful conversations. This work was supported in part by NSF grant 1205550 and the Fundamental Laws Initiative at Harvard.

\nopagebreak 

\bibliography{bunching}
\bibliographystyle{apsrev4-1}

\end{document}